\documentclass{article}

\usepackage{microtype}
\usepackage{graphicx}
\usepackage{subfig}
\usepackage{booktabs}

\usepackage{amsfonts}
\usepackage{bm}
\usepackage{mathtools}
\usepackage{paralist}

\usepackage{hyperref}
\usepackage{cleveref}
\crefname{equation}{}{}
\Crefname{equation}{}{}
\crefname{figure}{Figure}{Figures}
\crefname{section}{Section}{Sections}

\usepackage[accepted]{icml2021}

\newcommand{\norm}[1]{\lVert #1 \rVert}
\newcommand{\abs}[1]{\left|{#1}\right|}

\newcommand{\bdot}{{\color{black}.}}
\newcommand{\bcom}{{\color{black},}}
\newcommand{\Vc}{{\mathbf{c}}}
\newcommand{\Vf}{{\mathbf{f}}}
\newcommand{\Vh}{{\mathbf{h}}}
\newcommand{\Vu}{{\mathbf{u}}}
\newcommand{\Vv}{{\mathbf{v}}}
\newcommand{\Vw}{{\mathbf{w}}}
\newcommand{\Vx}{{\mathbf{x}}}
\newcommand{\VI}{{\mathbf{I}}}
\newcommand{\IR}{\mathbb{R}}
\newcommand{\IN}{\mathbb{N}}

\begin{document}
	
	\twocolumn[
	\icmltitle{Machine Learning for Auxiliary Sources}
	
	\begin{icmlauthorlist}
		\icmlauthor{Daniele Casati}{eth}
	\end{icmlauthorlist}
	
	\icmlaffiliation{eth}{Integrated Systems Laboratory, D-ITET, ETH Zurich, Switzerland}
	
	\icmlcorrespondingauthor{Daniele Casati}{dcasati@iis.ee.ethz.ch}
	
	\icmlkeywords{Method of auxiliary sources, Neural networks, Trefftz methods, Numerical simulation, Partial differential equations}
	
	\vskip 0.3in
	]
	
	\printAffiliationsAndNotice{}
	
	\begin{abstract}
			We rewrite the numerical ansatz of the Method of Auxiliary Sources (MAS), typically used in computational electromagnetics, as a neural network, i.e.\ as a composed function of linear and activation layers.
			MAS is a numerical method for Partial Differential Equations (PDEs) that employs point sources, which are also exact solutions of the considered PDE, as radial basis functions to match a given boundary condition.
			In the framework of neural networks we rely on optimization algorithms such as Adam to train MAS and find both its optimal coefficients and positions of the central singularities of the sources.
			In this work we also show that the MAS ansatz trained as a neural network can be used, in the case of an unknown function with a central singularity, to detect the position of such singularity.
	\end{abstract}

	\section{Introduction}
	\label{sec:introduction}
	
	\footnotetext{
		\emph{Abbreviations.}
		FEM: Finite Element Method.
		PDE: Partial Differential Equation.
		MMP: Multiple Multipole Program.
		MAS: Method of Auxiliary Sources.
	}

	\emph{Computational electromagnetics} studies how to numerically solve Maxwell's boundary value problems for engineering and scientific applications.
	An example is the simulation of \emph{plasmonic nanoparticles}, which can exhibit an interesting behavior (such as scattering) with electromagnetic radiation at a wavelength far larger than the particle size, depending on, e.g., the particle geometry \cite{Koch2018}.
	This phenomenon is relevant for many applications, such as solar cells or cancer treatment.

	For this kind of problems, several numerical methods have been developed based on a partition of the geometric domain: for example, the \emph{finite-difference time-domain} method \cite{KaneYee}, using a regular grid, or the \emph{Finite Element Method} (FEM) for frequency-domain Maxwell's equations \cite{hiptmair_2002}, using an unstructured mesh typically made of triangles (2D) or tetrahedra (3D).
	These approaches employ basis functions locally supported on the entities of the partition and therefore lead to large, sparse linear systems to solve.
	
	Conversely, another line of research in computational electromagnetics involves methods that do not need a mesh, as they make use of global basis functions (nonzero everywhere) that are exact solutions of the Partial Differential Equation (PDE) of interest.
	Given their nature, these basis functions do not need to be as many as the elements of FEM to achieve a good approximation, but, as we will see in a moment, require additional care.
	The numerical solution here is obtained by matching on a hypersurface either a boundary condition or interface conditions with another domain, discretized by
	\begin{inparaenum}[1)]
		\item
		different basis functions, as the PDE may be different there, or
		\item
		an entirely other method \cite{CASATI20191513}.
	\end{inparaenum}
	
	More details on these so-called \emph{Trefftz methods} \cite{HMP15} are given in the next section.
	Here it suffices to say that a common choice of Trefftz basis functions are \emph{point sources}, i.e.\ exact solutions that exhibit a central singularity.
	The centers of these singularities are placed in the complement of the respective domain of approximation, so that they are ignored by the computations.
	
	Furthermore, one would like to position these singularities in a way that the unknown function is well approximated by a linear combination of the sources.
	This holds true even if it can be proven that Trefftz methods enjoy exponential convergence when the unknown function has an analytic continuation beyond its approximation domain \cite{CASATI20191513}:
	as an example, please refer to the results in \cref{tab:L2error} below.
	In a way, the goal here is similar to choosing a high-quality unstructured mesh for FEM \cite{BRS2008} before assemblying and solving the related linear system.
	
	To find this ``optimal'' positioning of the sources, the state-of-the-art is based on elaborate heuristic rules developed over the years to support the user's manual positioning, especially in the context of computational electromagnetics, where one Trefftz method is the \emph{Multiple Multipole Program}\footnote{
	MMP is implemented by the open-source academic software \texttt{OpenMaXwell} \cite{HAFNER199921}, whose first development dates back to 1980 and which provides a graphical interface for the user to manually position the sources and check the corresponding numerical solution.} (MMP) \cite{HAFNER199921}.
	These heuristic rules are based on the \emph{curvature radius} \cite{Moreno:02} or on an \emph{unstructured surface mesh} \cite{Koch2018} of the hypersurface where the boundary condition needs to be imposed.
	Another line of research made use of \emph{genetic algorithms} \cite{1406223}.
	
	Here we propose an approach based on an optimization algorithm usually employed to train neural networks.
	As a numerical example for validation, let us consider for simplicity a Poisson's problem in a 2D bounded domain $\Omega$ -- in computational electromagnetics it can model, e.g., a vector magnetic potential orthogonal to the 2D plane -- with a manufactured solution of the type
	\begin{equation}
		\label{eq:solution}
		\rho_{x\sigma}^{-1}\sin(\theta_{x\sigma}) \bcom
	\end{equation}
	expressed in polar coordinates ($r\in[0,\infty)$, $\theta\in[0,2\pi)$) of $\Vx_{\sigma} \coloneqq \Vx - \bm{\sigma}$, with $\Vx,\bm{\sigma} \in \IR^2$ position vectors in Cartesian coordinates.
	Specifically, the center $\bm{\sigma}$ is taken in $\IR^2 \setminus \Omega$.

	Let us then approximate\footnote{
	More details on the approximation ansatz are given in \cref{sec:Trefftz}, on the experimental setup in \cref{sec:results}.} this problem with 3 point sources, randomly placed in $\IR^2 \setminus \Omega$ not too far from the hypersurface $\Gamma \coloneqq \partial\Omega$, and obtain their coefficients in a least-squares sense by matching the values of \cref{eq:solution} on selected points of $\Gamma$.
	
	\begin{table}[t]
		\caption{Relative $L^2(\Omega)$-error of 3 point sources to approximate \cref{eq:solution} for 10 different centers $\bm{\sigma}$ before and after the optimization ($5 \cdot 10^5$ epochs).}
		\label{tab:L2error}
		\vskip 0.15in
		\begin{center}
			\begin{small}
				\begin{sc}
					\begin{tabular}{p{0.20\columnwidth}p{0.20\columnwidth}}
						\toprule
						Initial ~~~~ positions & After ~~~~ algorithm \\
						\midrule
						51.19\% & 1.32\% \\
						32.07\% & 1.09\% \\
						44.83\% & 1.01\% \\
						10.70\% & 0.85\% \\
						78.28\% & 0.94\% \\
						15.37\% & 2.40\% \\
						28.23\% & 1.26\% \\
						32.15\% & 3.63\% \\
						10.11\% & 1.29\% \\
						\bottomrule
					\end{tabular}
				\end{sc}
			\end{small}
		\end{center}
		\vskip -0.1in
	\end{table}
	
	\Cref{tab:L2error} presents the corresponding $L^2(\Omega)$-error (given a bounded $\Omega$) of this ansatz with respect to \cref{eq:solution}, considering 10 different random centers $\bm{\sigma}$ not too far from $\Gamma$.
	The first column lists the relative errors (with respect to the $L^2(\Omega)$-norm of \cref{eq:solution}) for this random placement of the sources, which are as high as 78.28\%.
	Conversely, as shown by the errors in the second column, placing the sources with the optimization algorithm proposed in this work allows to achieve a better approximation with the same ansatz.

	\subsection{Summary and Structure}
	
	Compared to the state-of-the-art of Trefftz methods, by using the approach proposed in this paper we claim
	\begin{itemize}
		\item
		a higher accuracy, as shown by the results of \cref{tab:L2error}, which can also improve the more epochs are considered for the optimization,
		\item
		at a low additional runtime, given the efficient implementations of the optimization algorithms for neural networks: it takes a matter of seconds for the $5 \cdot 10^5$ epochs of \cref{tab:L2error}.
		\item
		This is also supported by the fact that the centers of the sources form a limited quantity of additional degrees of freedom for the optimization algorithm (on top of the coefficients of the sources).
		In fact, their number is not high (typically of an order of magnitude of 2) because of the exponential convergence of Trefftz methods.
		This allows to handle large system sizes for real-world engineering applications without particular restrictions.
	\end{itemize}

	The work is organized as follows: after this introduction, the fundamentals of Trefftz methods, specifically of MMP, are presented in \cref{sec:Trefftz}.
	Next, details on how to use the optimization algorithms of neural networks for MMP are given in \cref{sec:methodology}.
	This approach is then supported by the numerical results presented in \cref{sec:results}.
	Finally, \cref{sec:conclusions} concludes the paper.

	\section{Trefftz Methods}
	\label{sec:Trefftz}
	
	Trefftz methods employ exact solutions of the PDE as (global) basis functions.
	Hence, the main feature that characterizes a Trefftz method is its own discrete function space.
	As an example, for a 2D homogeneous Poisson's problem on a bounded domain $\Omega$, we work with the continuous \emph{Trefftz space} of functions
	\begin{equation}
		\label{eq:Trefftz}
		\mathcal{T}(\Omega) \coloneqq \Big\{ f \in H_\text{loc}^1(\Omega)\!:\: \nabla^2 v = 0 \Big\} \bdot
	\end{equation}	
	The functional form of the corresponding \emph{discrete} basis functions leads to different types of Trefftz methods:
	\begin{itemize}
		\item
		\emph{Plane waves} \cite{griffiths}
		or \emph{(generalized) harmonic polynomials} \cite{Moiola}
		constitute the most common choice \cite{HMP15}.
		\item
		If Trefftz basis functions solve an inhomogeneous problem, then we obtain the \emph{method of fundamental solutions} \cite{KUPRADZE196482}.
		\item
		Conversely, if they are point sources solving homogeneous equations (the right-hand side can be expressed by a known offset function), we get the \emph{Method of Auxiliary Sources} (MAS) \cite{UNK00}.
	\end{itemize}
	
	Concretely, from now on let us focus on a special case of MAS, i.e.\ the Multiple Multipole Program already introduced in \cref{sec:introduction}.
	
	The concept of this method was proposed by Ch.~Hafner in his dissertation \cite{HAF80bis} and popularized by his free code \texttt{OpenMaXwell} \cite{HAFNER199921} for 2D axisymmetric problems based on Maxwell's equations, especially in the fields of photonics and plasmonics\footnote{
	For example, one can consider the study of photonic structures presented in \cite{7465317} or plasmonic particles in \cite{Koch2018}.}.
	Hafner's MMP is in turn based on the much older work of G.~Mie and I.~N.~Vekua \cite{1900AnP...307..201M,Vekua}.
	Essentially, the Mie--Vekua approach expands some scalar field in a 2D multiply-connected domain \cite{Gamelin}
	by a \emph{multipole expansion} supplemented with generalized harmonic polynomials.
	Extending these ideas, MMP introduces more basis functions (\emph{multiple multipoles}) than required according to Vekua's theory \cite{Vekua} to span the Trefftz spaces \cref{eq:Trefftz}.	
	
	More specifically, \emph{multipoles} are potentials spawned by (anisotropic) point sources.
	These point sources are taken from the exact solutions of the homogeneous PDE, here Laplace's equation, which are subject to a condition at infinity when they are used to approximate the solution in an unbounded domain.

	A multipole can generally be written as $f(\Vx) \coloneqq g(\rho_{xc})\,h(\theta_{xc})$ or $\Vf(\Vx) \coloneqq g(\rho_{xc})\,\Vh(\theta_{xc},\varphi_{xc})$ in a polar/spherical coordinate system for $\Vx \in \IR^d$, $d=2,3$ ($r\in[0,\infty)$, $\theta\in[0,2\pi)$, $\varphi\in[0,\pi]$) with respect to its center $\Vc \in \IR^d$ ($\Vx,\Vc$ are position vectors in Cartesian coordinates).
	Here, $\left(\rho_{xc},\theta_{xc}\right)^\top$ and $\left(\rho_{xc},\theta_{xc},\varphi_{xc}\right)^\top$ are polar/spherical coordinates of the vector $\Vx_c \coloneqq \Vx - \Vc$.

	The radial dependence $g(\rho_{xc})$ has a center that presents a singularity, $\abs{g(\rho)}\to\infty$ for $\rho\to0$, and, possibly, the desired condition at infinity.
	Given the central singularity, multipoles are centered outside the domain in which they are used for approximation.

	On the other hand, the polar/spherical dependence $h$ or $\Vh$ is usually formulated in terms of trigonometric functions \cite{abramowitz+stegun}
	or \emph{(vector) spherical harmonics} \cite{0143-0807-12-4-007}.

	For the 2D Poisson's problem introduced in \cref{sec:introduction}, multipoles can have the form
	\begin{equation}
	\label{eq:expansion}
		\left(r,\theta\right) \mapsto
		\left\{
		\begin{aligned}
			&\log \rho_{xc}, && \\
			&\rho_{xc}^{-j}\cos(j\theta_{xc}), & j &= 1,\dots,\infty, \\
			&\rho_{xc}^{-j}\sin(j\theta_{xc}), & j &= 1,\dots,\infty,
		\end{aligned}
		\right.
	\end{equation}
	which also satisfy the condition at infinity\footnote{
	This condition is here unnecessary, as we work with a bounded $\Omega$ to compute $L^2(\Omega)$-errors for validation.}
	\begin{equation}
		\label{eq:decay}
		c\log\norm{\Vx} + \mathcal{O}(\norm{\Vx}^{-1}), \quad c\in\IR \bdot
	\end{equation}

\begin{figure}[ht]
		\vskip 0.2in
		\begin{center}
			\subfloat[$\log\rho$\label{fig:multipole00}]{%
				\includegraphics[width=0.31\columnwidth]{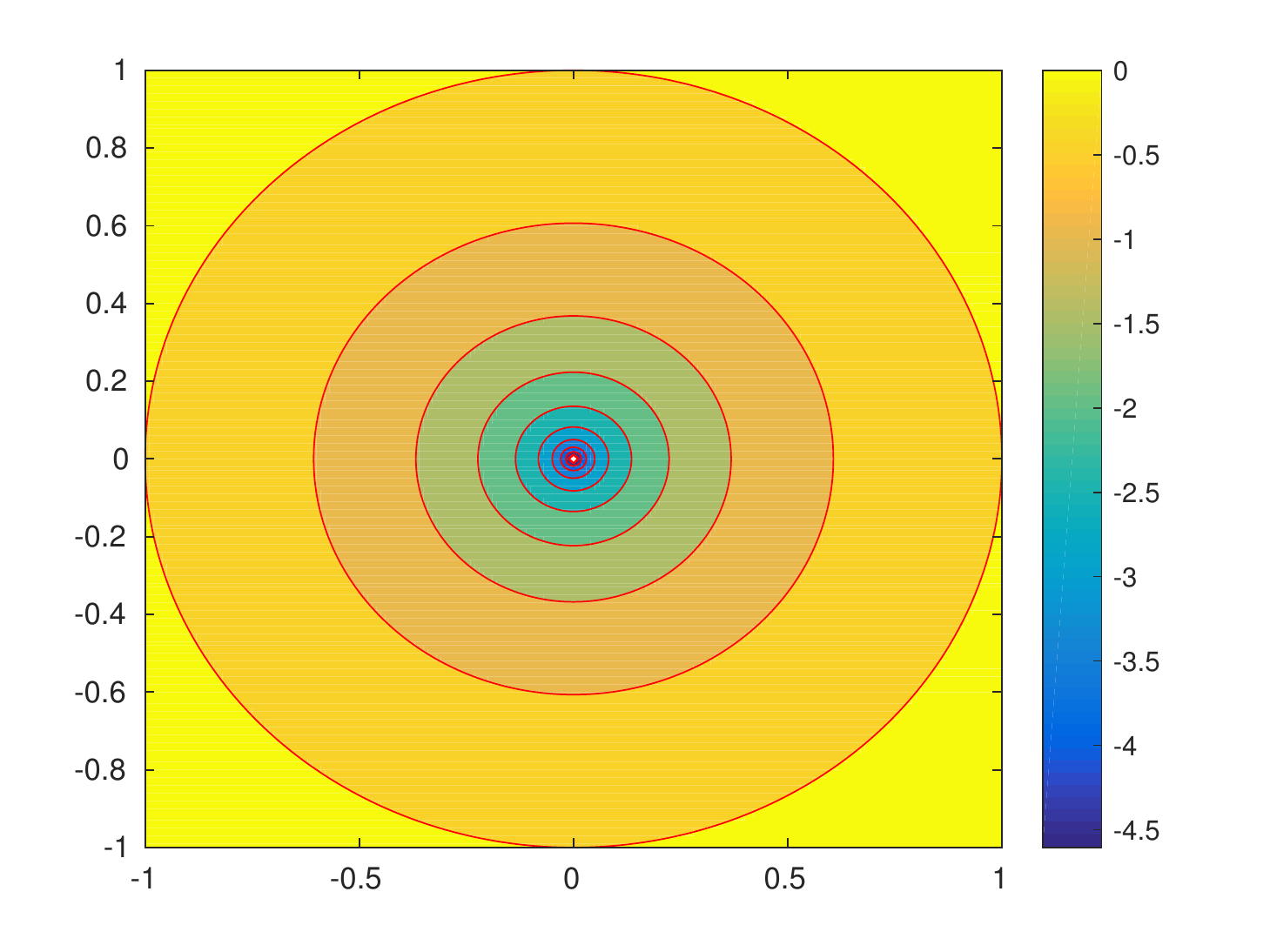}}
			\hfill
			\subfloat[$\rho^{-2}\cos(2\theta)$\label{fig:multipole20}]{%
				\includegraphics[width=0.31\columnwidth]{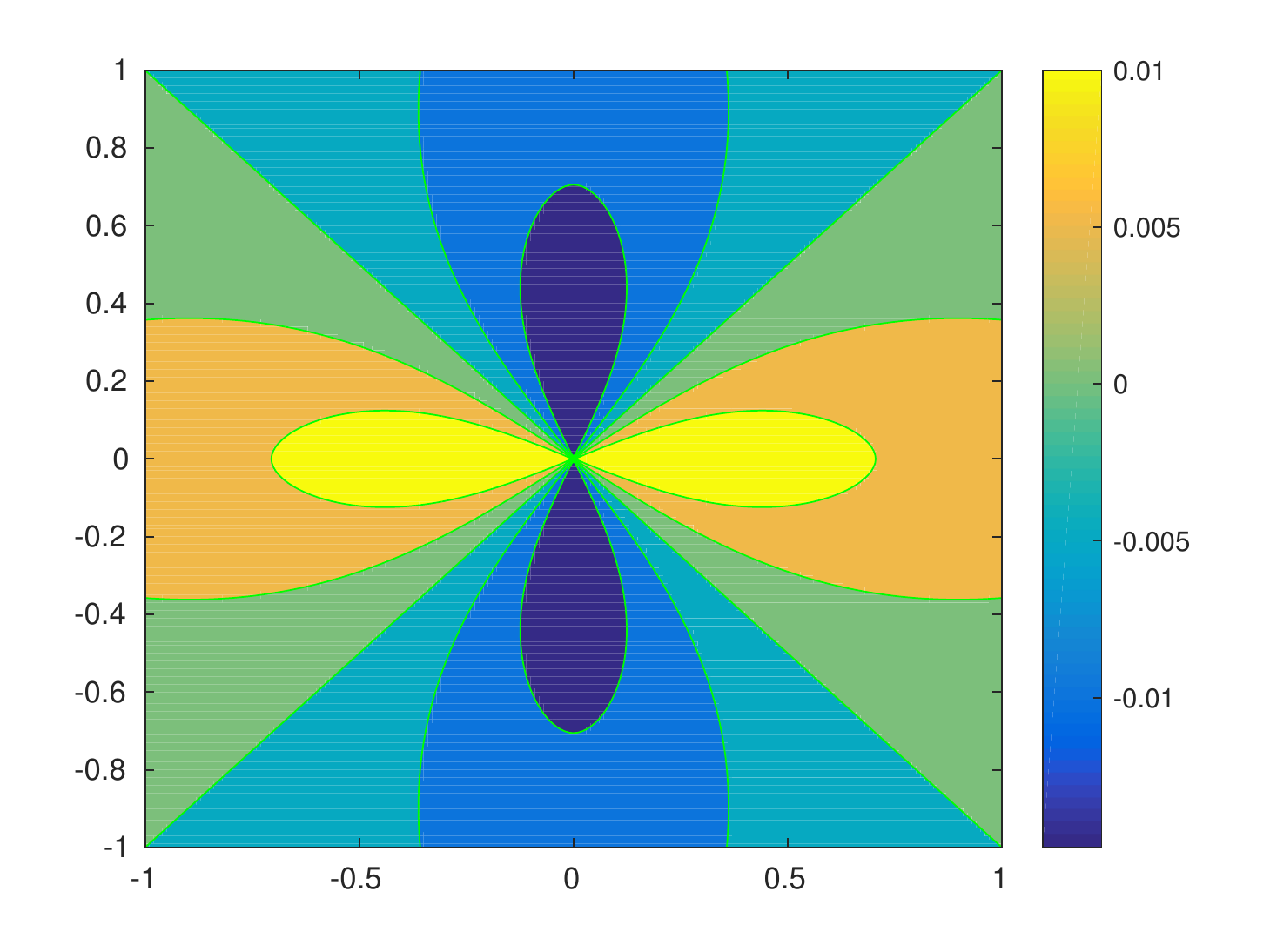}}
			\hfill
			\subfloat[$\rho^{-3}\cos(3\theta)$\label{fig:multipole30}]{%
				\includegraphics[width=0.31\columnwidth]{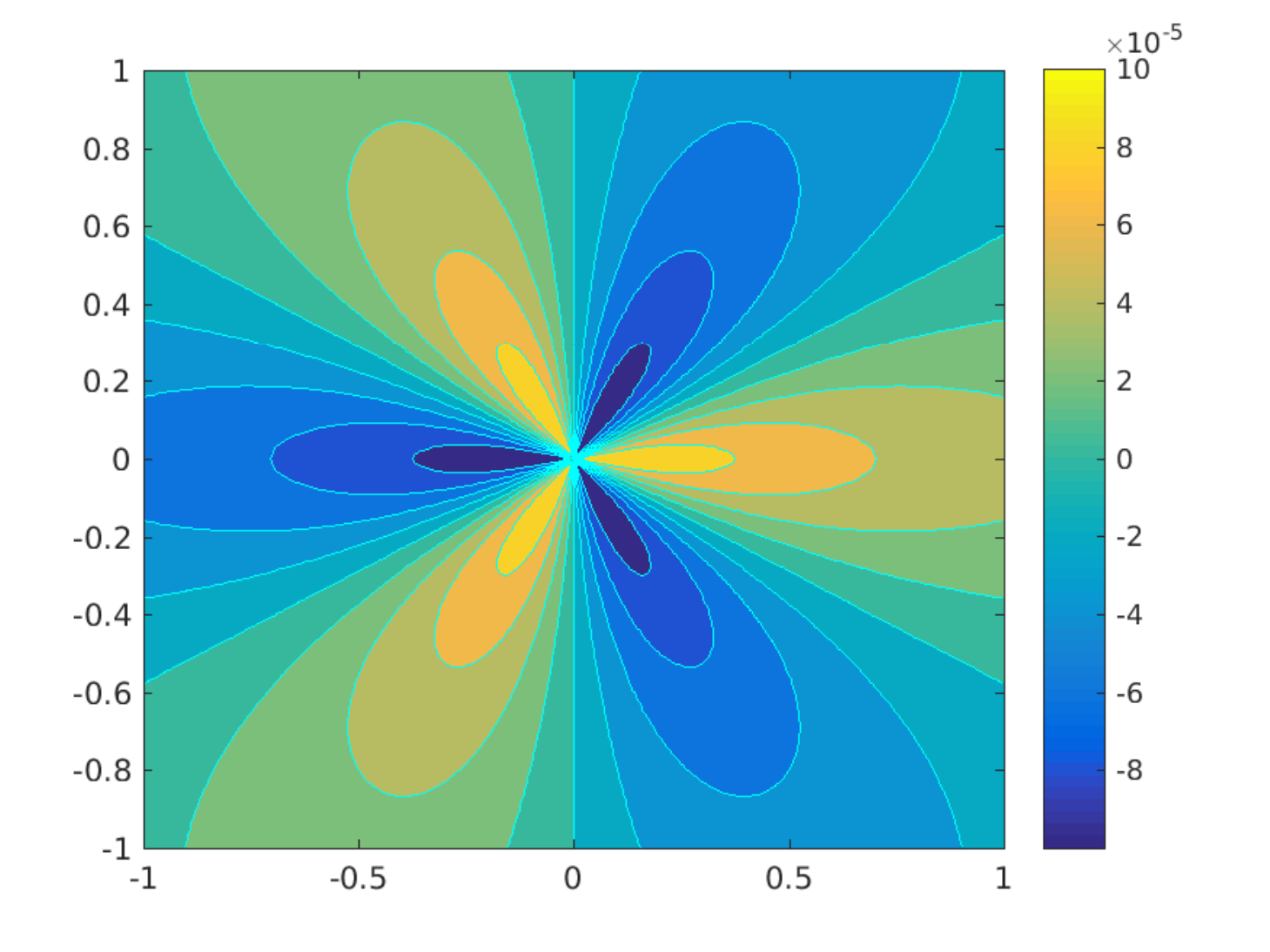}}
			\caption{Sample multipoles according to \cref{eq:expansion}, i.e.\ discrete basis functions of the MMP Trefftz space \cref{eq:Trefftz}.}
			\label{fig:multipoles}
		\end{center}
		\vskip -0.2in
	\end{figure}

	\Cref{fig:multipoles} shows three examples of multipoles according to \cref{eq:expansion} with center $\Vc = \mathbf{0}$.

	Each multipole from \cref{eq:expansion} is characterized by a location, i.e.\ its center $\Vc$, and the parameter $j$ (its degree), which can be assumed 0 for the case $\log \rho_{xc}$.
	When we place several multipoles at a given location up to a certain order $p$, which is the maximum degree of multipoles with that center, we use the term \emph{multipole expansion}.
	Summing the number of terms of all multipole expansions used for approximation (each with a different center) yields the total number of degrees of freedom of the discretized Trefftz space $\mathcal{T}^n(\Omega)$ from \cref{eq:Trefftz}.
	
	Once a discrete basis of multipoles has been chosen, there are several ways to find their coefficients such that the error with the boundary condition is minimized \cite{HMP15}: the most common is arguably \emph{collocation} on selected matching points of the hypersurface \cite{HAF80bis}, which aims at minimizing the $\ell^2$-error at the matching points in a least-squares sense.
	
	In this work we propose to use a \emph{gradient-based optimization algorithm}, typically employed to train neural networks that have many degrees of freedom, to optimize with respect to both the coefficients of the multipoles and the centers of their singularities.
	In other words, we do not preselect the centers and then find the corresponding optimal coefficients, in a similar way to precomputing a finite-element mesh, but we optimize both at the same time.
	This is doable because the number of centers scales logarithmically with respect to the number of matching points used for collocation, considering the exponential convergence of MMP \cite{CASATI20191513}.

	\section{MMP as a Neural Network}
	\label{sec:methodology}
	
	In this section we rewrite the numerical ansatz of MMP as a neural network, i.e.\ as a composed function of linear and activation layers.
	In this way, we are able to rely on efficient implementations of popular optimization algorithms for neural networks, such as the \emph{Adam algorithm} \cite{DBLP:journals/corr/KingmaB14}, as well as the \emph{automatic differentiation} \cite{DBLP:books/sp/Rall81} component of these implementations.
	
	The MMP ansatz approximates an unknown function as follows:
	\begin{equation}
		\label{eq:ansatz}
		\sum\limits_{i=1}^{n} \sum\limits_{j=0}^{p_i} w_{ij} f_{ij}( \Vx - \Vc_i ) \bcom
	\end{equation}
	where
	\begin{itemize}
		\item
		$n$ is the number of multipole expansions,
		\item
		$p_i$ is the order of the $i$-th multipole expansion, i.e.\ we consider terms in \cref{eq:expansion} from $j=0$ (corresponding to $\log \rho_{xc}$) to $p_i$, and
		\item
		$w_{ij} \in \IR$ is the coefficient of $f_{ij}( \Vx - \Vc_i ) \in \mathcal{T}^n(\Omega)$, with $\Vx,\Vc_i \in \IR^2$ and $\Vc_i$ center of the $i$-th multipole expansion.
	\end{itemize}

	Following the formalism of neural networks \cite{Haykin}, we can rewrite \cref{eq:ansatz} as a 3-layer neural network, given $\Vx \in \IR^2$ as the \emph{input variable}:
	\begin{enumerate}
		\item
		The first \emph{linear layer} is represented by the following affine transformation with a nonzero shift:
		\begin{equation}
			\Vx \mapsto
			\begin{pmatrix}
				\left.
				\begin{array}{c}
					\VI_{2,2} \\
					\vdots \\
					\VI_{2,2}
				\end{array}
				\right\} p_1 \\
				\vdots \\
				\left.
				\begin{array}{c}
					\VI_{2,2} \\
					\vdots \\
					\VI_{2,2}
				\end{array}
				\right\} p_n
			\end{pmatrix}
			\Vx
			-
			\begin{pmatrix}
				\left.
				\begin{array}{c}
					\Vc_1 \\
					\vdots \\
					\Vc_1
				\end{array}
				\right\} p_1 \\
				\vdots \\
				\left.
				\begin{array}{c}
					\Vc_n \\
					\vdots \\
					\Vc_n
				\end{array}
				\right\} p_n
			\end{pmatrix} = \Vu \in \IR^{2 m} \bcom
		\end{equation}
		where $m \coloneqq \sum_{i=1}^n (p_i+1)$ and the shift (\emph{bias}) is made of the centers of the multipoles, to be determined by the optimization.
		\item
		The \emph{activation layer} is composed of several \emph{``many-to-one'' activation functions}, as they map pairs of variables to a single one:
		\begin{equation}
			\Vf(\Vu) = \Vv \in \IR^{m} \bcom
		\end{equation}
		where each activation function $f_{ij}: \IR^2 \to \IR$, $j=0,\dots,p_i$, $i=1,\dots,n$, is a multipole.
		Examples of ``many-to-one'' activation functions from the literature of neural networks include \emph{softmax}, \emph{max pooling}, \emph{maxout}, and \emph{gating} \cite{DBLP:journals/corr/abs-1710-05941}.
		\item
		Finally, the third layer is linear without bias:
		\begin{equation}
			\Vw^\top \cdot \Vv = y \in \IR \bcom
		\end{equation}
		where the weights $\Vw \in \IR^{m}$ are the coefficients of the multipole expansions.
	\end{enumerate}
	
\begin{figure}[ht]
		\vskip 0.2in
		\begin{center}
			\centerline{\includegraphics[width=\columnwidth]{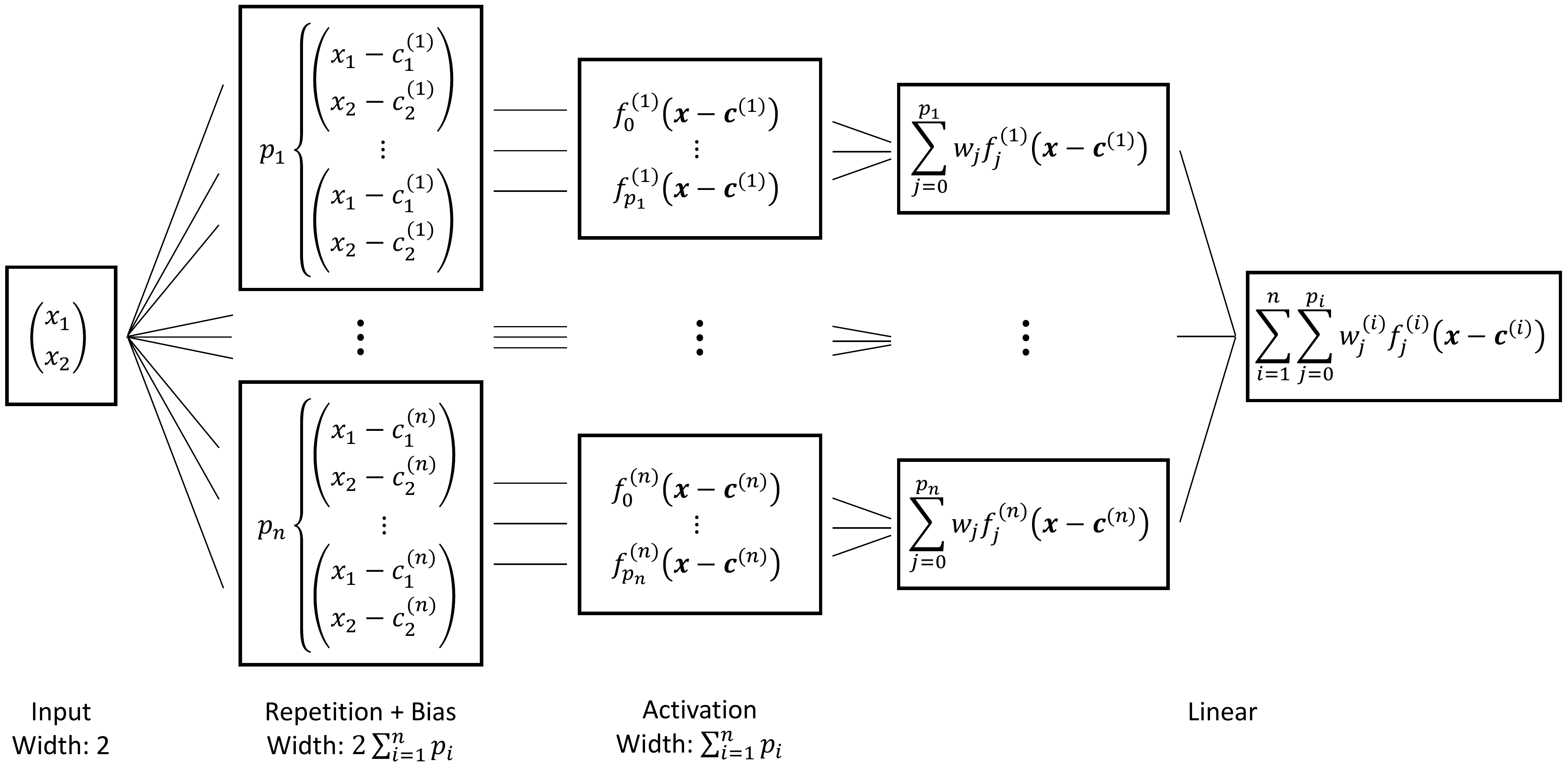}}
			\caption{Neural network representation of the MMP ansatz.}
			\label{fig:network}
		\end{center}
		\vskip -0.2in
	\end{figure}

	\Cref{fig:network} schematizes the neural network representation of the MMP ansatz described above.
	
	Based on this representation, one can see that the centers of singularities $\Vc_i$, $i=1,\dots,n$, are, together with the coefficients $w_{ij}$, $j=0,\dots,p_i$, degrees of freedom of this neural network.
	The total number of degrees of freedom is $2n + \sum_{i=1}^n (p_i+1)$: note that, in case of few high-order multipole expansions, the additional degrees of freedom constituted by their centers ($2n$) do not have much impact on the total number.

	For the loss function of this neural network we follow the collocation method, whose goal is to minimize the $\ell^2(\Gamma)$-error between the MMP ansatz \cref{eq:ansatz} and the boundary condition on the chosen matching points:
	\begin{equation}
		L(\Vc,\Vw) \coloneqq \sum\limits_{l=1}^{N} \left[ \sum\limits_{i=1}^{n} \sum\limits_{j=0}^{p_i} w_{ij} f_{ij}( \Vx_l - \Vc_i ) - y_l \right]^2 \bcom
	\end{equation}
	where $y_l$, $l=1,\dots,N$, are the evaluations of the unknown (assuming a Dirichlet boundary condition) on the $N$ matching points $\Vx_l$.

	\section{Numerical Results}
	\label{sec:results}
	
	As bounded domain $\Omega$ for a 2D Poisson's problem we choose the interior region of a flower-shaped curve $\Gamma \coloneqq \partial\Omega$, parameterized by the formula
	\begin{equation}
		\label{eq:flower}
		\left( R(\theta)\cos\theta, R(\theta)\sin\theta \right)^\top, \quad R(\theta) = \alpha(\beta + \gamma\cos(K\theta))
	\end{equation}
	in Cartesian coordinates, with $\theta\in[0,2\pi)$, $\alpha,\beta,\gamma \in \IR$, and $K\in\IN$.
	We set $\alpha=0.5$, $\beta=1$, $\gamma=0.5$, and $K=5$, and choose $N=100$ points of $\Gamma$ from equidistant values of $\theta\in[0,2\pi)$ to serve as matching points.
	
\begin{figure}[ht]
		\vskip 0.2in
		\begin{center}
			\subfloat[Flower-shaped curve according to \cref{eq:flower} (100 matching points).\label{fig:flower_R}]{%
				\includegraphics[width=0.49\columnwidth]{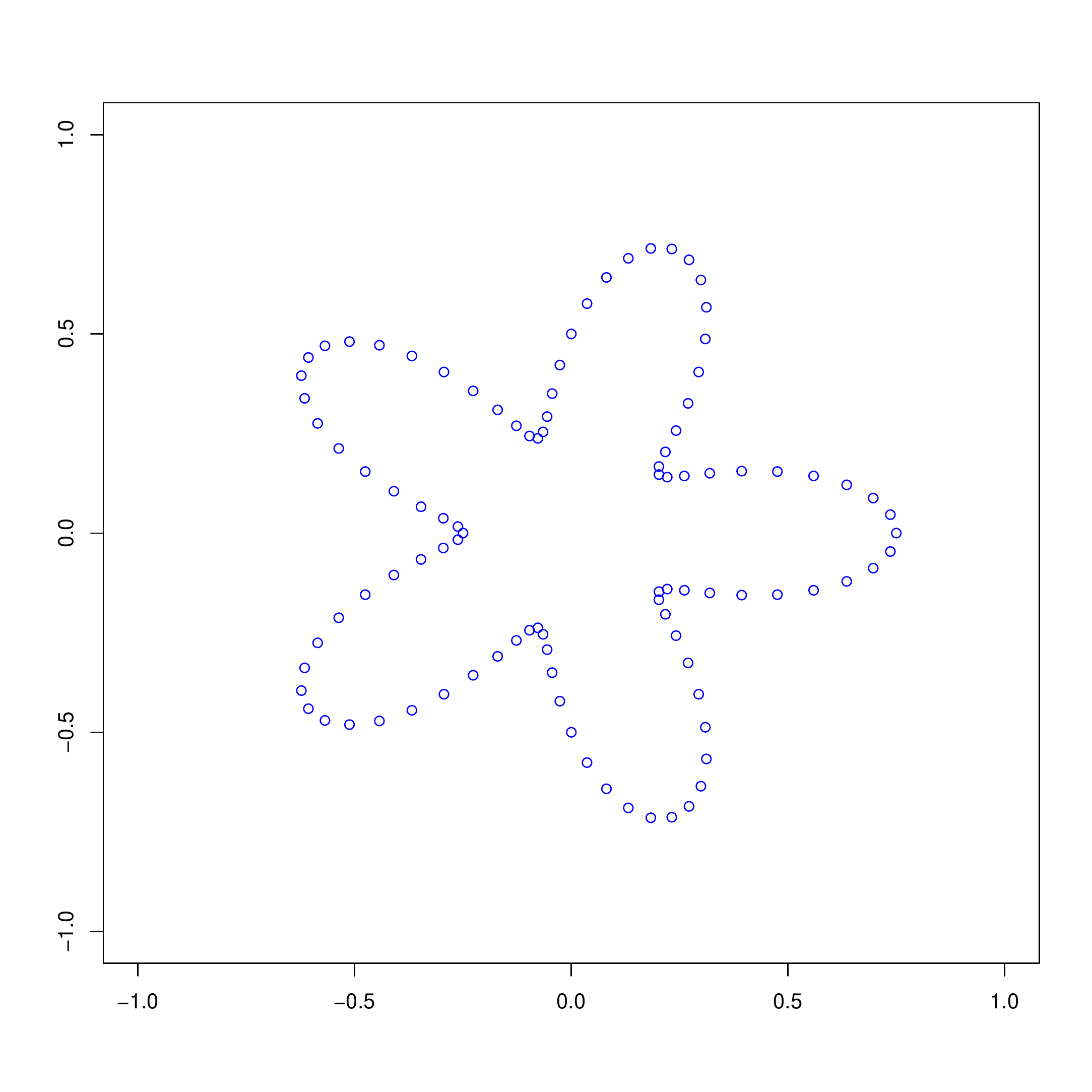}}
			\hfill
			\subfloat[Sample mesh of the flower-shaped domain.\label{fig:flower_netgen}]{%
				\includegraphics[width=0.44\columnwidth]{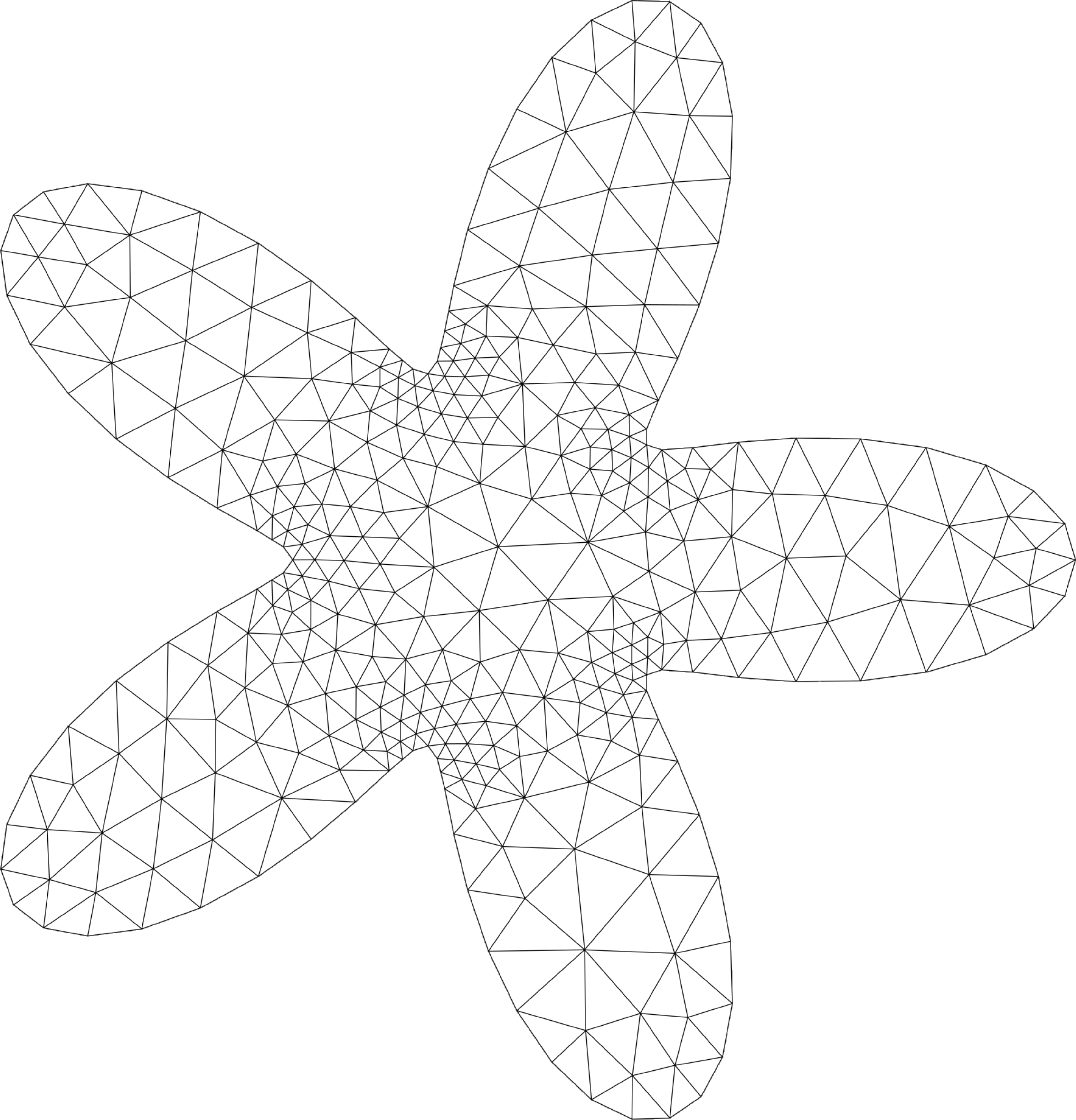}}
			\caption{Flower-shaped curve and sample mesh.}
			\label{fig:flower}
		\end{center}
		\vskip -0.2in
	\end{figure}

	\Cref{fig:flower_R} shows the flower-shaped curve according to \cref{eq:flower}, \cref{fig:flower_netgen} the meshed domain with triangles.
	This mesh is used to compute the $L^2(\Omega)$-error of the MMP approximation with respect to manufactured solutions of type \cref{eq:solution} (see \cref{sec:introduction}) with different centers $\bm{\sigma} \in \IR^2 \setminus \Omega$.
	For the numerical quadrature of the error, we employ the Gaussian quadrature rule of order 5 (polynomials up to the 5-th order are integrated exactly) on triangles.

	In the following we discuss two experiments solving this boundary value problem with MMP:
	first, we investigate the dependence of the Adam optimization on the number of multipole expansions used for approximation (\cref{subsec:results_number}), then on the order of a single multipole expansion (\cref{subsec:results_order}).

	\subsection{Dependence on the number of multipole expansions}
	\label{subsec:results_number}
	
	With the Adam algorithm we find the coefficients and centers of singularities of the MMP expansions that minimize the boundary $\ell^2(\Gamma)$-error with a manufactured solution on selected matching points.
	The ``dataset'' to train the MMP ansatz as a neural network is made of coordinates of matching points as input observations and the corresponding evaluations of the manufactured solution as output.
	
\begin{figure}[ht]
		\vskip 0.2in
		\begin{center}
			\subfloat[In red, 10 initial centers of multipole expansions.\label{fig:flower_R_circle}]{%
				\includegraphics[width=0.49\linewidth]{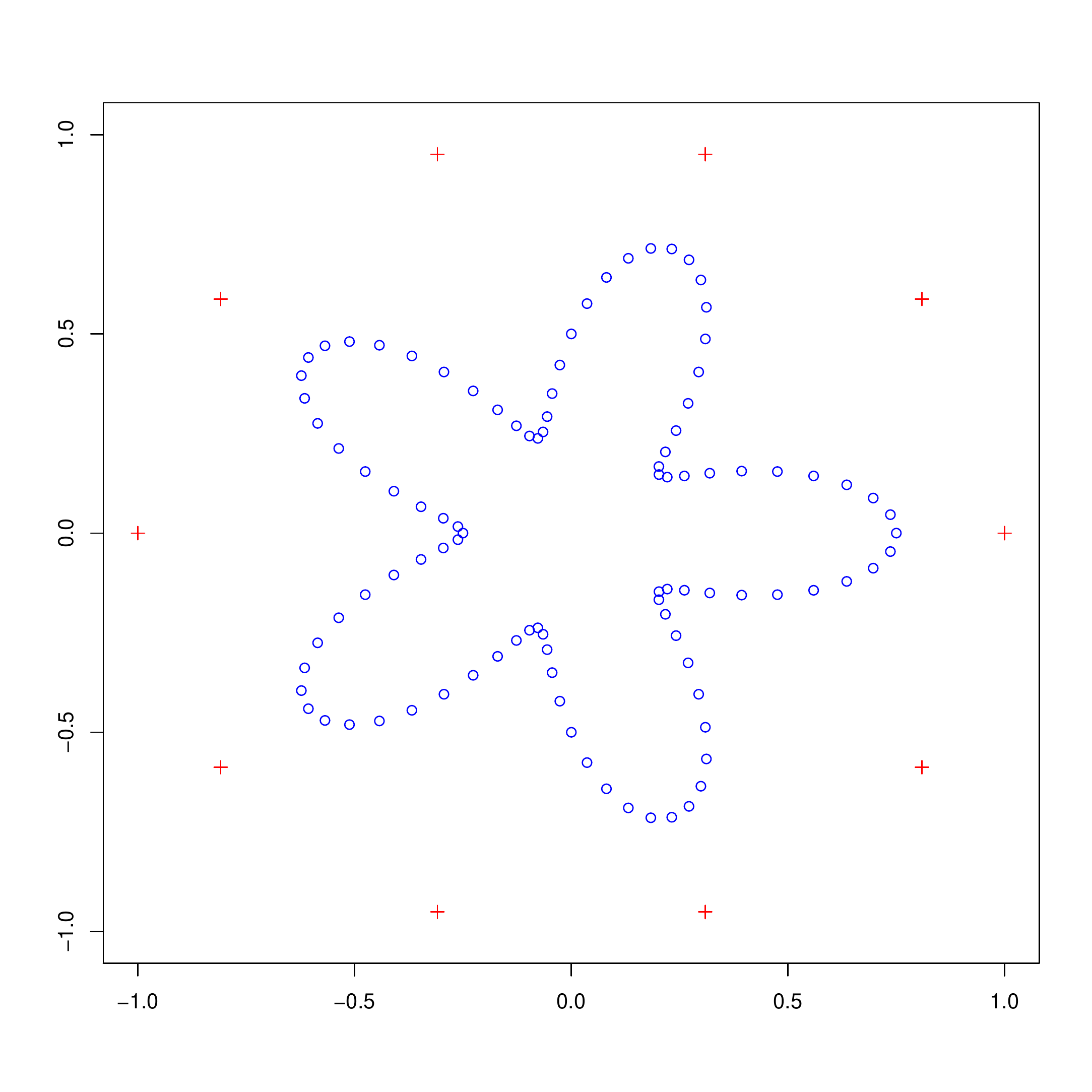}}
			\hfill
			\subfloat[In red, 100 centers of manufactured solutions.\label{fig:flower_R_flower}]{%
				\includegraphics[width=0.49\linewidth]{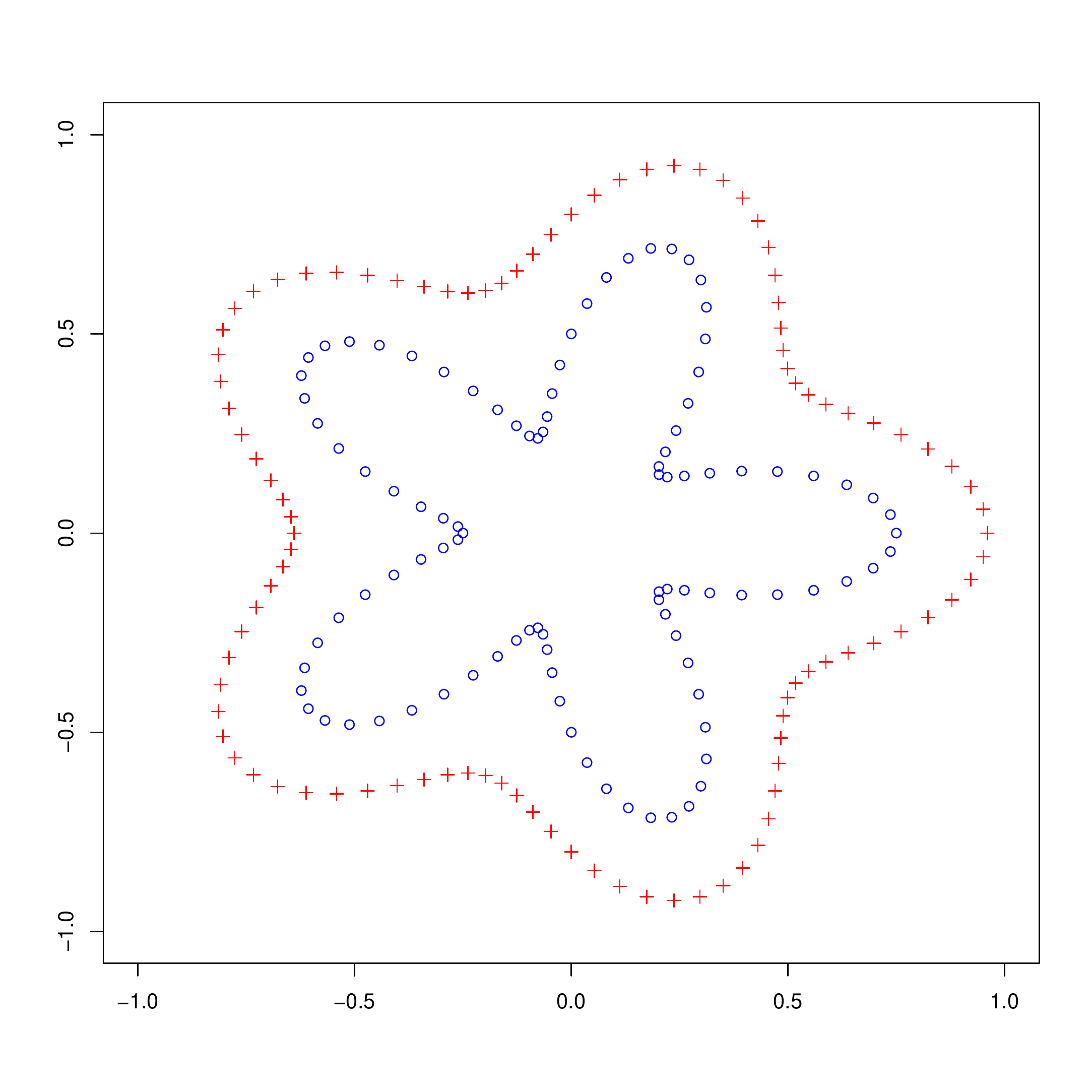}}
			\caption{Initial centers of multipoles and centers of manufactured solutions.}
			\label{fig:flower_R_}
		\end{center}
		\vskip -0.2in
	\end{figure}

	To approximate \cref{eq:solution}, we choose an MMP ansatz made of several multipole expansions (we vary their number $n$), each of order 1.
	Before the Adam optimization, their centers are initially disposed on the unit circle, which is external to the bounded domain $\Omega$ of \cref{fig:flower} (see \cref{fig:flower_R_circle}).
	The corresponding initial values for the multipole coefficients are obtained from the collocation method, given these centers.
	
	The Adam algorithm is then run for a number of epochs until the training loss becomes smaller than 0.05, for at most $5 \cdot 10^5$ epochs.
	The learning rate is 0.1 and the batch size is the full dataset (100 matching points): this is justified because the number of observations is equal to the number of matching points, set by the user, which must therefore be humanely manageable (here 100).
	
	We perform this procedure for 100 manufactured solutions, each centered on an equidistant point of the curve \cref{eq:flower} with parameters $\alpha=0.8$, $\beta=1$, $\gamma=0.2$, and $K=5$ (see \cref{fig:flower_R_flower}).

\begin{figure}[ht]
		\vskip 0.2in
		\begin{center}
			\centerline{\includegraphics[width=\columnwidth]{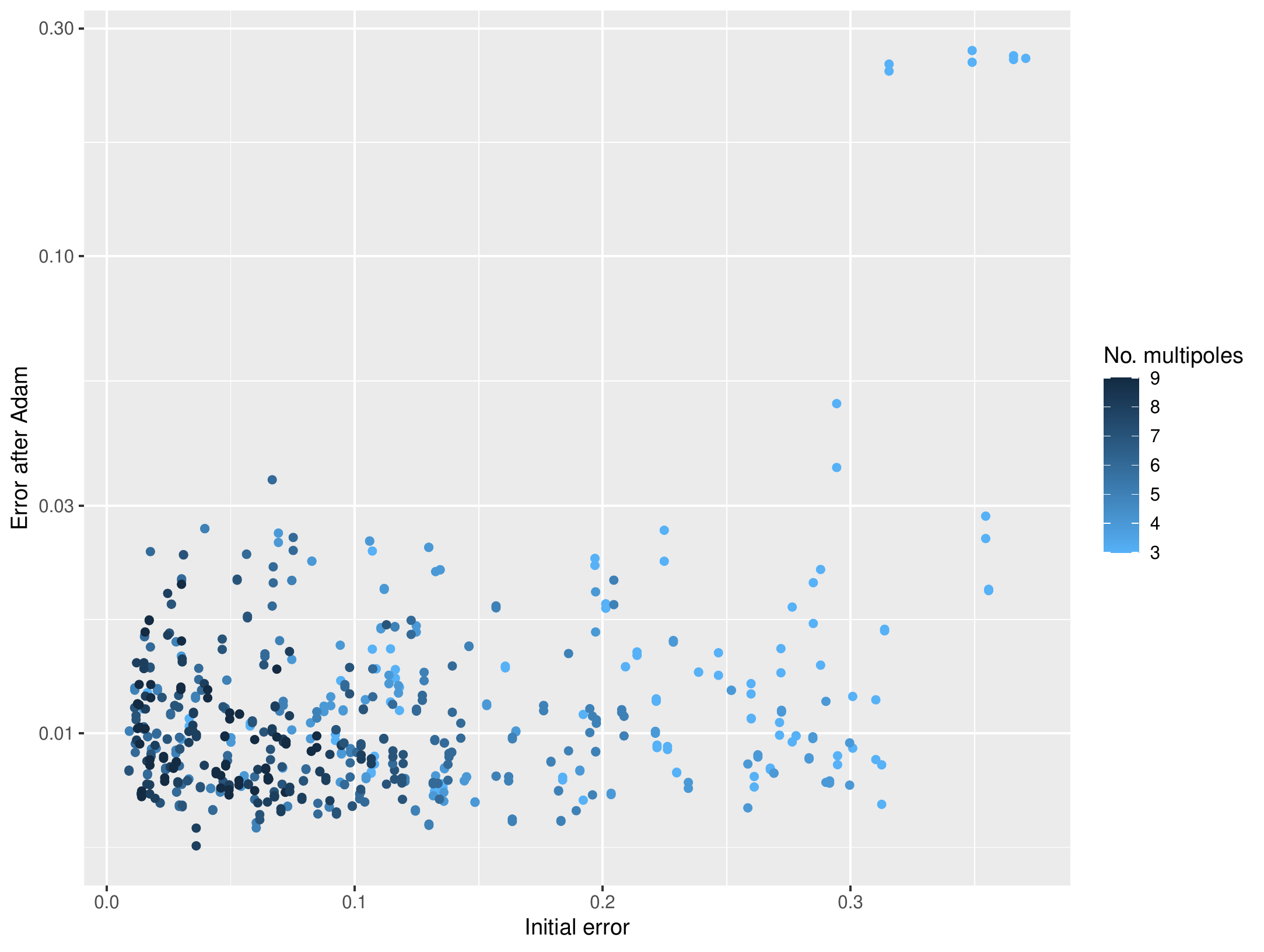}}
			\caption{$L^2(\Omega)$-error before and after the Adam optimization (many multipole expansions with order $1$).}
			\label{fig:test_1}
		\end{center}
		\vskip -0.2in
	\end{figure}

	\Cref{fig:test_1} shows the $L^2(\Omega)$-error (normalized with the $L^2(\Omega)$-norm of the manufactured solution) produced by the initial positioning on the unit circle\footnote{
	As proven in \cite{SAK16},
	if the solution \cref{eq:solution} of the 2D Poisson's equation possesses an analytic extension beyond $\Omega$, specifically into the region of $\IR^2 \setminus \Omega$ between $\Gamma$ and the curve $\Sigma$ along which the multipole expansions are placed, then we expect exponential convergence in terms of the number of multipole expansions (or their orders).
	
	This is however \emph{not} the case of \cref{fig:flower_R_circle}, which shows the unit circle $\Sigma$ in red, given the solutions with the centers shown in \cref{fig:flower_R_circle}.
	In this way, the proposed approach is tested for an initial positioning that does not present exponential convergence when the number of multipole expansions is increased.} (see \cref{fig:flower_R_circle}) and after the Adam optimization.
	The number of multipole expansions is in the range $[3,9]$ to investigate its impact on the optimization result.

	Notice that, while the initial relative error shows considerable variation, the final error tends to stabilize at a median value of $\approx$ 0.98\%.
	At the same time, there are a few cases where the optimization fails and the final error is noticeably larger than the initial one: they take place when $5 \cdot 10^5$ epochs are not enough for the loss to become smaller than 0.05.
	This happens in 2.17\% of observations, especially for high numbers of considered expansions, as there are more degrees of freedom to optimize.
	Furthermore, 101 observations (out of 700) are excluded from \cref{fig:test_1} because the optimization stopped at the first step: the training loss was already smaller than 0.05 with the initial positioning of the multipoles.

\begin{figure}[ht]
		\vskip 0.2in
		\begin{center}
			\centerline{\includegraphics[width=\columnwidth]{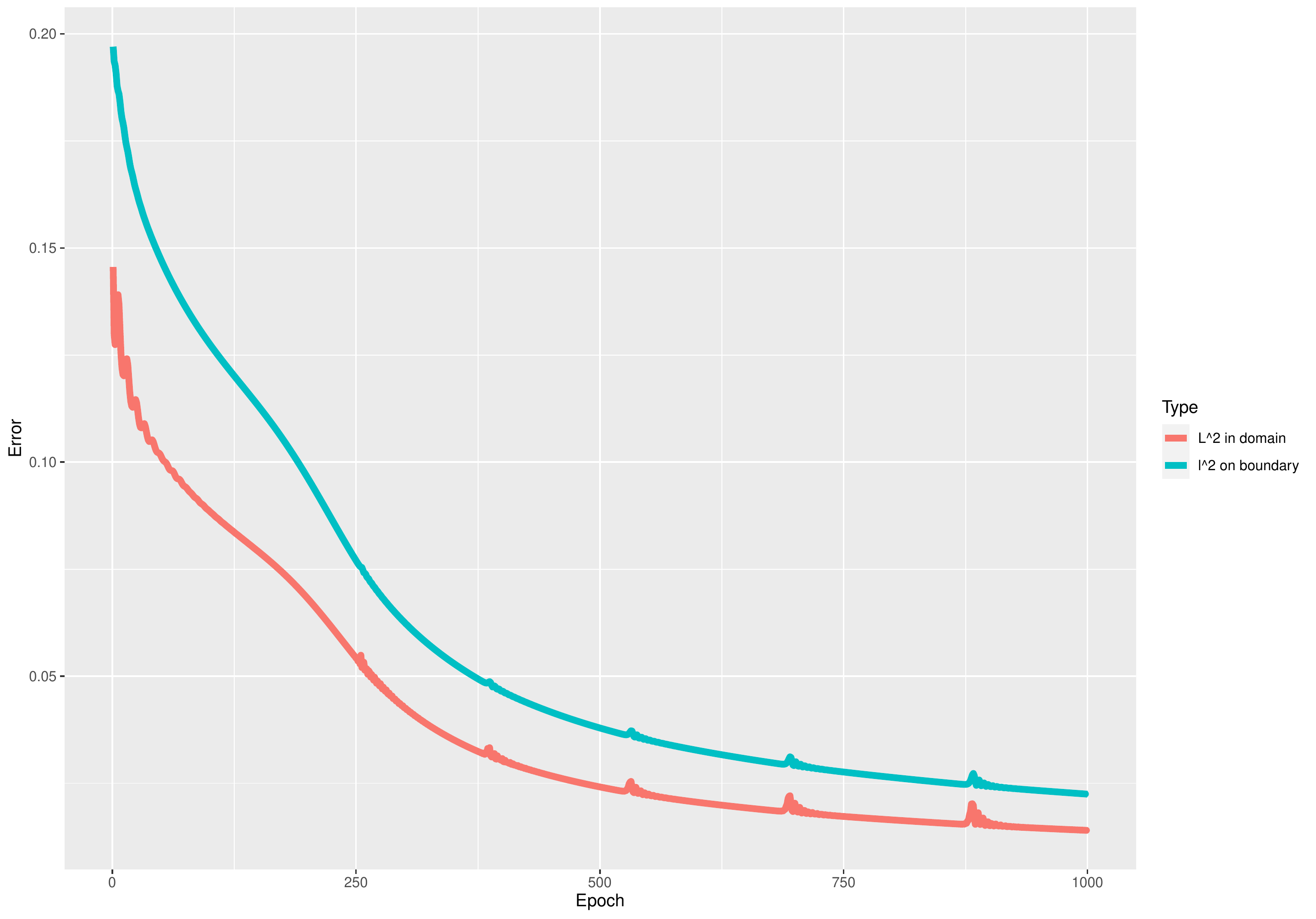}}
			\caption{Training loss (i.e.\ $\ell^2(\Gamma)$-error) and $L^2(\Omega)$-error over 1\,000 Adam epochs for a sample manufactured solution.}
			\label{fig:epoch}
		\end{center}
		\vskip -0.2in
	\end{figure}

	For a sample manufactured solution and 3 multipole expansions, \cref{fig:epoch} reports the decay of the training loss of the Adam algorithm (the $\ell^2(\Gamma)$-error on matching points) over 1\,000 epochs and the corresponding normalized $L^2(\Omega)$-error.
	The decay of the training loss is not monotone, even if we take the maximal batch size during the training process, i.e.\ the full dataset (so the gradient always points in the right direction), because of local minima.
	The $L^2(\Omega)$-error presents more pronounced spikes, as we are not optimizing with respect to this value.

	\subsection{Dependence on the order of one multipole expansion}
	\label{subsec:results_order}
	
	Instead of multiple multipole expansions, let us now train one expansion (one center) of a given order (which we vary).
	As initial position for the center we choose a random point outside $\Omega$ and inside the square centered in the origin with side length 4, while the initial coefficients of the expansion are obtained from the collocation method for this center.
	
	The Adam algorithm is then run for a number of epochs until the training loss becomes smaller than 0.05, for at most $5 \cdot 10^5$ epochs.
	The learning rate is 0.1 and the batch size is the full dataset.
	
	We perform this procedure for 100 manufactured solutions, each centered on an equidistant point of the curve \cref{eq:flower} with parameters $\alpha=0.8$, $\beta=1$, $\gamma=0.2$, and $N=5$ (see \cref{fig:flower_R_flower}).
	
\begin{figure}[ht]
		\vskip 0.2in
		\begin{center}
			\centerline{\includegraphics[width=\columnwidth]{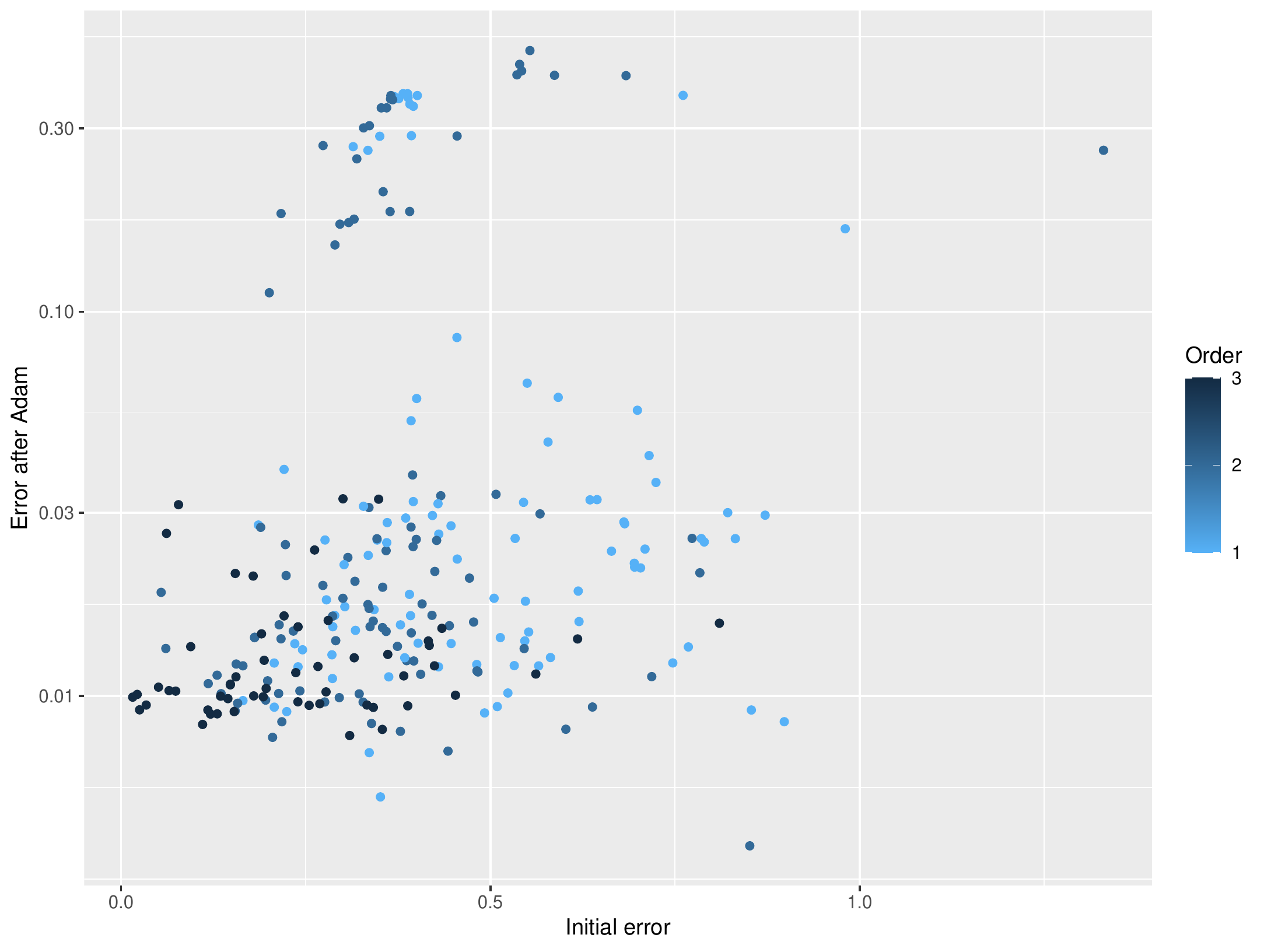}}
			\caption{$L^2(\Omega)$-error before and after the Adam optimization (one multipole expansion with different orders).}
			\label{fig:test_2}
		\end{center}
		\vskip -0.2in
	\end{figure}

	\Cref{fig:test_2} shows the normalized $L^2(\Omega)$-error produced by the initial position and after the Adam optimization.
	The order of the multipole expansion is in the range $[1,3]$.
	
	Similarly to \cref{fig:test_1}, while the initial error shows considerable variation, the final relative error tends to stabilize at a median value of $\approx$ 1.62\%.
	Contrarily to \cref{fig:test_1}, for no observation the final error is larger than the initial one.
	
	Moreover, a single multipole expansion could also be used to find the center of the singularity of \cref{eq:solution} itself.
	In fact, the best position for the multipole expansion to approximate \cref{eq:solution} should be at the singularity itself.
	This could be proven by contradiction by taking a circle with radius $\rho$ and considering the $\ell^2(\Gamma)$-error for a multipole expansion centered at a point of this circle.
	This error will depend on $\rho$ and become minimal when $\rho=0$.
	
\begin{figure}[ht]
		\vskip 0.2in
		\begin{center}
			\centerline{\includegraphics[width=\columnwidth]{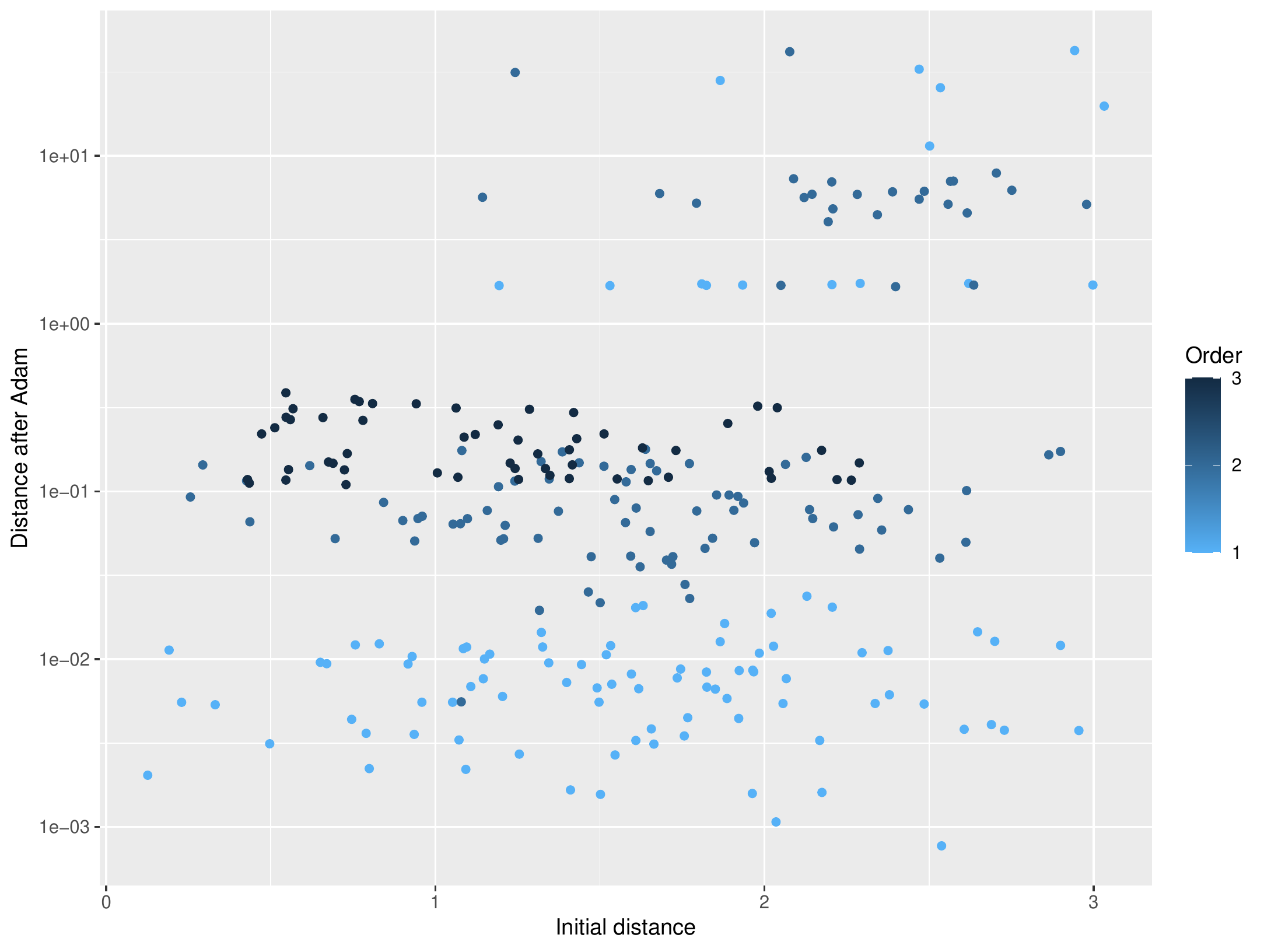}}
			\caption{$\ell^2$-distance between solution and expansion before and after the Adam optimization (one multipole expansion with different orders).}
			\label{fig:distance}
		\end{center}
		\vskip -0.2in
	\end{figure}

	\Cref{fig:distance} shows the $\ell^2$-distance between the singularity of the solution and the center of one multipole expansion before and after the Adam optimization.
	In 12.16\% of observations the initial distance is smaller than the one after the Adam optimization: however, in all these cases the training loss did not become smaller than 0.05 in $5 \cdot 10^5$ epochs.

	\subsection{Implementation}
	\label{subsec:implementation}
	
	Our code is written in \texttt{Python~3}.
	
	For meshing and numerical integration, we rely on \texttt{Netgen/NGSolve} \cite{NGSolve}.
	For the Adam optimization, we rely on \texttt{PyTorch} \cite{PyTorch}.
	By defining a new \texttt{torch.nn.Module} for the first and second layer of the MMP ansatz as a neural network (\cref{sec:methodology}), we can use the automatic differentiation tool of \texttt{PyTorch} for the Jacobians needed by the backpropagation step of the Adam algorithm.
	Furthermore, we exploit the \texttt{PyTorch} parallelization on GPUs when training each neural network and the \texttt{Python} multiprocessing module to parallelize over the manufactured solutions.

	\section{Conclusions}
	\label{sec:conclusions}
	
	We have shown that gradient-based optimization algorithms commonly used to train neural networks, such as the Adam algorithm, can help with overcoming a flaw of MAS, namely the heuristics needed to place its point sources, by optimizing with respect to these positions (together with the coefficients of the point sources).
	
	Future work will involve
	\begin{inparaenum}[1)]
		\item
		applying this approach to other problems, i.e.\ with different boundaries, manufactured solutions, and PDEs, and
		\item
		using a genetic algorithm to optimize also with respect to the number and orders of the multipole expansions (total number of degrees of freedom of MMP).
	\end{inparaenum}
	These are the metaparameters of the MMP ansatz as a neural network, which now have to be chosen by the user.
	A too large number of degrees of freedom for MMP should be penalized by the genetic procedure.
	
	\bibliography{literature}
	\bibliographystyle{icml2021}
	
\end{document}